\documentclass[lettersize,journal]{IEEEtran}
\usepackage{amsmath,amsfonts}
\usepackage{algorithmic}
\usepackage{algorithm}
\usepackage{array}
\usepackage[caption=false,font=small,labelfont=sf,textfont=sf]{subfig}
\usepackage{textcomp}
\usepackage{stfloats}
\usepackage{url}
\usepackage{verbatim}
\usepackage{graphicx}
\usepackage{color}
\usepackage[dvipsnames]{xcolor}
\usepackage{soul}
\usepackage{cite}
\graphicspath{{Figures/}}

\usepackage{amssymb}
\usepackage{accents}

\DeclareMathOperator{\Ln}{Ln}

\usepackage{hyperref}  
\usepackage{cleveref}
\crefname{figure}{Fig.}{Figs.}
\Crefname{figure}{Fig.}{Figs.}
\crefname{equation}{Eq.}{Eqs.}
\Crefname{equation}{Eq.}{Eqs.}
\usepackage{import}
\usepackage{booktabs}

\begin{document}

\title{Instantaneous Complex Phase and Frequency: Conceptual Clarification and Equivalence between Formulations}

\author{César García-Veloso,~\IEEEmembership{Member,~IEEE}, Mario Paolone,~\IEEEmembership{Fellow,~IEEE}, Federico Milano,~\IEEEmembership{Fellow,~IEEE}
\thanks{César García-Veloso and Mario Paolone are with the Distributed Electrical Systems Laboratory, École Polytechnique Fédérale de Lausanne, 1015 Lausanne, Switzerland (e-mail: cesar.garciaveloso@epfl.ch; mario.paolone@epfl.ch).}%
\thanks{Federico Milano is with the School of Electrical and Electronic Engineering, University College Dublin, Belfield, D04V1W8 Dublin, Ireland (e-mail: federico.milano@ieee.org).}}



\maketitle
\begin{abstract}
This letter seeks to clarify the different existing definitions of both instantaneous complex phase and frequency as well as their equivalence under standard modeling assumptions considered for transmission systems, i.e. balanced positive sequence operation, sole presence of electro-mechanical transient dynamics and absence of harmonics and interharmonics.  To achieve this, the two fundamental definitions, i.e., those based on either the use of (i) analytic signals or (ii) space vectors, together with the premises used for their formulation, are presented and their relationship shown. Lastly, a unified notation and terminology to avoid confusion is proposed.
\end{abstract}


\section{Introduction}
Recently, a proliferation of works has emerged e.g. \cite{Mountevelis2024_Con_Cont_CFC,  Enrich2024_Dyn_CFC_GFC, Quan2025_CFLL, Milano2025_Local_synch} leveraging the re-introduction done  in \cite{Milano2022_Complex_Frequency} of the concept of \textit{instantaneous complex frequency} (ICF) within the field of power systems based on the usage of the Park vector. However, the notion of ICF was already formulated many decades ago in the works of both Linden \cite{Linden1958_ICF} and Hahn \cite{Hahn1959_ICF} using the concept of analytic signals. The latter of the two even applied it to electric circuits as far back as 1964 \cite{Hahn1964_Complex_Freq}. Moreover, as also pointed by Hahn himself in \cite{HT_Trans_App_Hand2010}, time-independent complex frequency is a well-established concept in the fields of signal processing and systems theory as it represents the independent variable of the Laplace transform.
Historically, as also indicated by Hahn in \cite{HT_Trans_App_Hand2010} and exemplified in \cite{Boashash1991_Est_int_IF}, a significant amount of confusion and misunderstandings have surrounded the concept of instantaneous frequency (IF). The reason is no other but the adoption of the same label, IF, by different authors, to accommodate different definitions and the subsequent pointless argumentation regarding which is one correct. Similarly, together with \cite{Milano2022_Complex_Frequency}, a definition of \textit{instantaneous complex phase} (ICP) has also been proposed in \cite{Lei2022_Complex_Angle_PLL} based on the Clarke space vector with respect to three-phase phase-locked loops (PLLs). 
Thus, the purpose of this letter is to prevent an analogous situation to that seen for the IF to further develop around the concepts of both ICF and ICP. This is done by: (i) showcasing the different currently proposed formulations as well as how and under which conditions these relate to each other and (ii) proposing an unified notation and terminology to avoid confusion.

\section{Equivalence between Existing Formulations}
\subsection{Analytic Signals: Hahn \cite{Hahn1964_Complex_Freq, Hahn2003_Uniqueness_A_ph_analytic, HT_Trans_App_Hand2010}}
Hahn's definitions of ICP and ICF of a time domain signal $v(t)$ rely on the use of the Gabor complex signal or analytic signal $\tilde{v}(t)$:
\begin{equation}
\tilde{v}(t) = v(t)+\jmath\mathcal{H}\{v(t)\} = u(t)e^{\jmath\theta (t)} , \label{eq:ana_sig_def}
\end{equation} 
where $\mathcal{H}$ denotes the Hilbert Transform (HT) \cite{HT_Trans_App_Hand2010}, $u(t)$ the instantaneous amplitude and $\theta (t)$ the instantaneous phase. The ICP $\bar{\phi}(t)$  is then defined as\footnote{The same notation is consistently used for the ICP and ICF, with different sub-indexes employed for each respective formulation, i.e. $\mathfrak{h}$ for Hahn, $\mathfrak{m}$ for Milano and $\mathfrak{l}$ for Lei.}:
\begin{equation}
\bar{\phi}_{\mathfrak{h}}(t) = \Ln{(\tilde{v}(t))} = \Ln{(u(t))}+\jmath\theta (t) \, , \label{eq:hahn_complex_phase}
\end{equation}
where $\Ln$ is used to highlight the multi-branch nature of the natural logarithm of complex numbers. The ICF $\bar{s}(t)$ is subsequently obtained as the time derivative of the ICP:
\begin{equation}
\bar{s}_{\mathfrak{h}}(t) = \dot{\bar{\phi}}_{\mathfrak{h}}(t) = \frac{\dot{u}(t)}{u(t)}+ \jmath\dot{\theta} (t) = \varrho_{\mathfrak{h}}(t) +\jmath\omega_{\mathfrak{h}}(t) \, ,
\label{eq:hahn_complex_freq}
\end{equation}
where $\varrho_{\mathfrak{h}}(t)$ and $\omega_{\mathfrak{h}}(t)$ refer respectively to the instantaneous radial and angular frequencies \cite{HT_Trans_App_Hand2010}. This definition is universally valid as it does not entail any assumption nor consideration for the signal $v(t)$ besides those required for the existence of its HT, i.e. $v(t)$ must have finite-energy (square-integrable) or be absolutely integrable. Furthermore, as proven in \cite{Hahn2003_Uniqueness_A_ph_analytic}, this formulation ensures the ICF is uniquely defined. Regarding its use within three-phase power systems, it can be applied to any generic voltage or current signal for each individual phase.
\subsection{Space Vectors: Milano \cite{Milano2022_Complex_Frequency} and Lei \cite{Lei2022_Complex_Angle_PLL}}
The definition of the ICF proposed by Milano \cite{Milano2022_Complex_Frequency}  --- called \textit{complex frequency} in this reference --- is based on the use of the Park vector $\bar{v}_{dq}(t)$ and consequently specifically applies to three-phase systems:
\begin{equation}
\bar{v}_{dq}(t) = v_d(t)+\jmath v_q(t) = u_{dq}(t)e^{\jmath\vartheta (t)}  ,\label{eq:park_vec_def}
\end{equation}
where $v_d(t)$ and $v_q(t)$ represent the direct and quadrature components and $u_{dq}(t)$ and $\vartheta (t)$ the instantaneous magnitude and phase.  Note that the Park vector is defined based on the $dq$-axis components of the Park transform and fully represents a three-phase system under balanced conditions and in the absence of homopolar components. Although \cite{Milano2022_Complex_Frequency} only explicitly defines ICF, both the ICP\footnote{Despite (\ref{eq:fed_comp_ph}), i.e. \cite[Eq. (29)]{Milano2022_Complex_Frequency}, not being explicitly referred to as the ICP in \cite{Milano2022_Complex_Frequency}, the ICF is indeed shown to be its time derivative. Moreover, the term is explicitly introduced as such based on Milano's ICF in \cite[Eq. (4)]{Buttner2025_Complex_Phase}.} and ICF correspond to:
\begin{equation}
\bar{\phi}_{\mathfrak{m}}(t) =  \ln{(u_{dq}}(t))+ \jmath\vartheta (t) \, ,\label{eq:fed_comp_ph}
\end{equation}
\begin{equation}
\bar{s}_{\mathfrak{m}}(t) = \dot{\bar{\phi}}_{\mathfrak{m}}(t) = \frac{d(\ln{(u_{dq}}(t)))}{dt}+ \jmath\dot{\vartheta} (t) = \varrho_{\mathfrak{m}}(t) +\jmath\omega_{\mathfrak{m}}(t) \, . \label{eq:fed_comp_freq}
\end{equation}
It is important to clarify that, while balanced positive sequence operating conditions, absence of harmonics and interharmonics as well as electro-mechanical transient dynamics are required to link (\ref{eq:fed_comp_freq}) to the time derivative of the complex powers, as shown in \cite{Milano2022_Complex_Frequency}, (\ref{eq:fed_comp_ph}-\ref{eq:fed_comp_freq}) are intrinsically generic and only require the application of the Park transform. 

Let us denote with $\delta_{dq}(t)$ the rotation angle of the Park $dq$-axis reference frame. As both Clarke and Park consider the same analysis frames --- the only difference being that $\delta_{dq}(t) = 0$ in the Clarke transform ---, the definition proposed by Lei \cite{Lei2022_Complex_Angle_PLL} for the ICP, as well as its associated ICF, based on the Clarke space vector, $\bar{v}_{\alpha\beta}(t)$, can be considered analogous to that of \cite{Milano2022_Complex_Frequency}. Indeed, an obvious equivalence can easily be established between the definitions of ICP and ICF presented by \cite{Lei2022_Complex_Angle_PLL} and \cite{Milano2022_Complex_Frequency} as:
\begin{subequations}\label{eq:fed_lei_rel}
\begin{equation}
    \bar{\phi}_{\mathfrak{m}}(t)  = \jmath ( \bar{\phi}_{\mathfrak{l}}(t) - \delta_{dq}(t)) \, ,
\end{equation}
\begin{equation}
    \bar{s}_{\mathfrak{m}}(t)  = \jmath (\bar{s}_{\mathfrak{l}}(t) -\omega_{dq}(t) ) \, ,
\end{equation}
\end{subequations}
where $\omega_{dq}(t) = \dot{\delta}_{dq}(t)$.  Note that Lei's definition permutes real and imaginary parts compared to Milano's \cite{Milano2022_Complex_Frequency}. In view of geometrical considerations which interpret $\varrho$ as a dilatation and $\omega$ as a rotation (see Section \ref{sec:proposed}), the notation in \cite{Milano2022_Complex_Frequency} is formally more consistent.

\subsection{Hahn - Milano Relationship}\label{sec:relationship}
Consider a generic single-tone three-phase power system signal $\pmb{v}_{abc}(t)$:
\begin{equation}\label{eq:gen_3Ph_sig}
\pmb{v}_{abc}(t) = \begin{bmatrix}
    v_a(t)\\v_{b}(t)\\v_{c}(t)
\end{bmatrix} , \;\! \begin{array}{c}
     v_{k}(t) = u_{k}(t)\cos{(\theta_{k}(t)+\zeta_{k}(t))} \, , \\
     \theta_{k}(t) = \omega_o t +\varphi_{k}(t) \, , \\
     k \in\{a,b,c\} \, ,
\end{array}  
\end{equation} 
where $k$ is used to represent each individual phase, $\zeta_k(t)$ the angular shifts between phases and $u_{k}(t)$ and $\theta_{k}(t)$ the time varying magnitude and phase for which $\omega_o$ denotes the constant nominal system frequency and $\varphi_{k}(t)$ its time-dependent part. It holds that $\bar{v}_{\alpha\beta}(t)$ contains both positive and negative sequence spectral components according to:
\begin{equation}\label{eq:park_lyon}
\bar{v}_{\alpha\beta}(t) \propto \bar{v}_p(t) = 1/2 (\underaccent{\bar}{V}_p(t)e^{\jmath\omega_o t} + \underaccent{\bar}{V}^{*}_n(t)e^{-\jmath\omega_o t}) \, ,
\end{equation}
where $\bar{v}_p(t)$ is the Lyon positive sequence vector \cite{Paap2000_Symmetrical_Components}, $\underaccent{\bar}{V}_p(t)$ and $\underaccent{\bar}{V}_n(t)$ are the positive and negative sequence phasors and $^{*}$ represents the complex conjugate operation.

On the other hand, the analytic signal of each phase $k$ can be calculated as:
\begin{equation}
\tilde{v}_k(t) = v_k(t)+\jmath\mathcal{H}\{v_k(t)\} \simeq u_k(t)e^{\jmath(\theta_k (t)+\zeta_{k}(t))},  \label{eq:ana_sig_def_ps}
\end{equation}
where $\simeq$ in (\ref{eq:ana_sig_def_ps}) is used to symbolize that, according to the Bedrosian theorem \cite{Bedrosian1963_BT}, this equality only holds if there is no spectral overlap between $u_k(t)$ and $\theta_k(t)+ \zeta_{k}(t)$. However, as shown in \cite{Dervisakdic2020_Beyond_Phasors,Karpilow2025_FBA_FPGA}, this is indeed the case for the slower dynamics that characterize electro-mechanical transients as those considered in \cite{Milano2022_Complex_Frequency} to link (\ref{eq:fed_comp_freq}) with the time derivatives of complex powers.
Lastly, under positive sequence balanced conditions, and by taking $k=a$ and $\zeta_{a}(t) = 0$ for simplicity it follows that the spectrum of $\bar{v}_{\alpha\beta}(t)$ matches that of $\tilde{v}_{a}(t) $\cite{Ferrero1991Comp3PhNonSinusoidal}:
\begin{equation}\label{eq:3Ph_bal_pos}
\begin{cases}
    \underaccent{\bar}{V}_p(t) = \underaccent{\bar}{V}_a(t)\\ 
    \underaccent{\bar}{V}_n^{*}(t) = 0
\end{cases}\!\!\! \Rightarrow \bar{v}_p(t) =  \frac{1}{2} \tilde{v}_{a}(t) = \frac{1}{2}\bar{v}_{\alpha\beta}(t) \, .
\end{equation}
Lastly, the Park and Clarke vectors can be related simply by the rotation angle $\delta_{dq}(t)$ applied to the Park frame as:
\begin{equation}\label{eq:Clarke_Park}
\bar{v}_{dq}(t) = \bar{v}_{\alpha\beta}(t) e^{-\jmath\delta_{dq}(t)} =  \tilde{v}_{a}(t)e^{-\jmath\delta_{dq}(t)} \, .
\end{equation}
Therefore, the relationships between ICP and ICF of Hahn and Milano are respectively given by:
\begin{subequations}\label{eq:ICP_ICF_Hahn_Milano}
\begin{equation}\label{eq:ICP_Hahn_Milano}
\bar{\phi}_{\mathfrak{m}}(t) = \bar{\phi}_{\mathfrak{h}}(t) -\jmath \delta_{dq}(t) \, ,
\end{equation}
\begin{equation}\label{eq:ICF_Hahn_Milano}
\bar{s}_{\mathfrak{m}}(t) = \bar{s}_{\mathfrak{h}}(t) -\jmath \omega_{dq}(t) \, .
\end{equation}
\end{subequations}
Note that the selection of any other phase, i.e. $k \in \{b,c\}$, under balanced conditions will simply introduce an additional phase offset in the imaginary part.

\section{Proposed Terminology}\label{sec:proposed}
The equivalence between ICP and ICF of Hahn and Milano given in (\ref{eq:ICP_ICF_Hahn_Milano}) holds only in specific conditions, namely, balanced positive sequence, sole presence of electro-mechanical transient dynamics and absence of harmonics and interharmonics\footnote{The latter is required so that $\bar{v}_{\alpha\beta}(t)$ is an analytic signal, i.e. $v_{\beta}(t) = \mathcal{H}\{v_{\alpha}(t)\}$. Specifically, given the spectral content of $\bar{v}_{\alpha\beta}(t)$ \cite{Ferrero1991Comp3PhNonSinusoidal}, the absence of negative and zero sequence harmonics as well as interharmonics is needed.}. Although these conditions are not always met (e.g. in power distribution systems or, in general, in the presence of electromagnetic transients), they are used in standard modeling of power transmission systems. Therefore, the proven equivalence constitutes an important clarification for those involved in the modeling and analysis of such systems. In general, however, Hahn's and Milano's definitions refer to fundamentally unrelated concepts and the use of the same terminology is not recommended.
\begin{figure}[!t]
\centering
\scriptsize
\includegraphics[width=\linewidth]{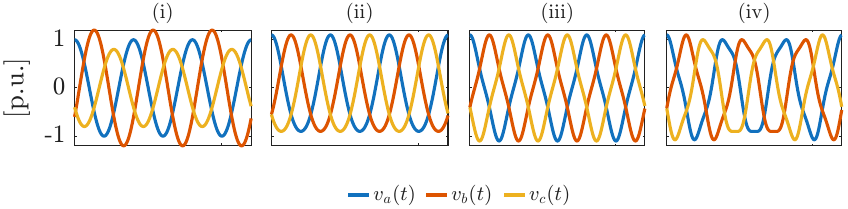} (a)\vspace{1mm}
\includegraphics[width=\linewidth]{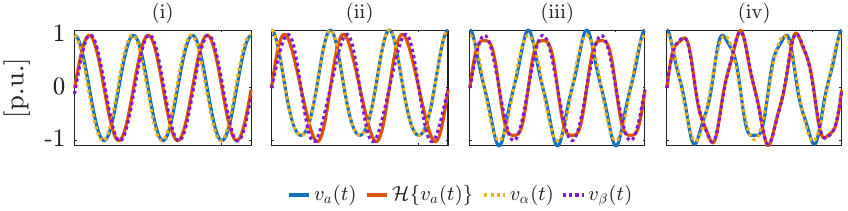} (b)\vspace{1mm}
\includegraphics[width=\linewidth]{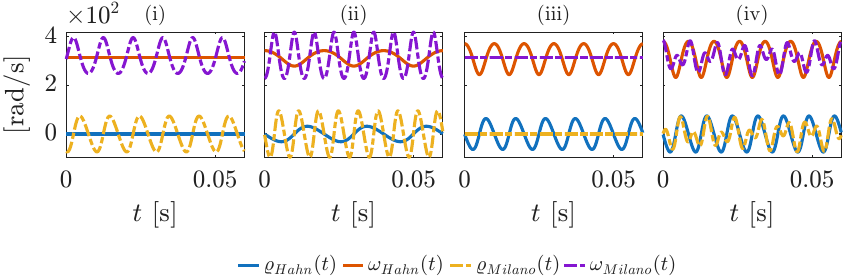} (c)
\caption{Sample cases where equivalence conditions are not met. Each column represents the superposition of: (i) Fundamental amplitude unbalance, (ii) a 2\textsuperscript{nd} harmonic, (iii) a 3\textsuperscript{rd} harmonic and (iv) an interharmonic. Each row depicts: (a) Phase voltages, (b) components of the analytic signal of phase $a$ as well as $\alpha$ and $\beta$ components of the Clarke vector and (c) ICF according to Hahn's and Milano's definitions. For the latter $\delta_{dq}(t) = 0$ is considered.}
\label{fig_ICF_IPF}
\end{figure}
An illustrative example is provided in \Cref{fig_ICF_IPF} where Hahn's and Milano's ICF are shown across four different cases where the equivalence conditions are not met. It depicts the phase voltages (\Cref{fig_ICF_IPF}(a)) together with the components of the analytic signal and the $\alpha$ and $\beta$ components of the Clarke vector (\Cref{fig_ICF_IPF}(b)) as well as the corresponding ICF according to Hahn's and Milano's definitions (\Cref{fig_ICF_IPF}(c)). Note that for the latter $\delta_{dq}(t) = 0$ has been considered for simplicity.

Since ICP and ICF were first defined for the analytic signal and represent a generic and field-agnostic case, it is the authors' view that such notation should be retained.  Moreover, as highlighted by Hahn \cite{Hahn2003_Uniqueness_A_ph_analytic}, the scientific and engineering communities already use the term IF to refer to rate of change of the phase of an analytic signal. 

On the other hand, as it is more recent and has not been yet widely adopted by the power system community, it appears easier and more consistent to change the notation of the ICF defined in \cite{Milano2022_Complex_Frequency}.  With this aim, it is relevant to note that the complex frequency defined in \cite{Milano2022_Complex_Frequency} can be viewed as a special case of the \textit{geometric frequency} defined in \cite{Milano2022_Geometrical_Freq}.  This quantity is, in the most general case, a multivector composed of a scalar and a bivector, as defined in the Clifford algebra $Cl_{1,2}$ for three-dimensional spaces, which corresponds to the Pauli algebra --- see, for example, \cite{Jancewicz:1989}. 

The starting point of the geometric approach is the vector of voltages $\pmb{v}_{abc}(t)$ at a point of a three-phase circuit.  This voltage is assumed to be the velocity vector field of a generalized position, which, in this analogy, has the dimension of a magnetic flux.  Then, the geometric frequency is defined as:
\begin{equation}
    \label{eq:geom_freq}
    \hat{\pmb{\Omega}}(t) = \frac{\pmb{v}_{abc}(t) \cdot \dot{\pmb{v}}_{abc}(t)}{|\pmb{v}_{abc}(t)|^2} + \frac{\pmb{v}_{abc}(t) \wedge \dot{\pmb{v}}_{abc}(t)}{|\pmb{v}_{abc}(t)|^2} \, ,
\end{equation}
where $\cdot$ and $\wedge$ indicate inner and outer products, respectively, and $|\pmb{x}| = \sqrt{\pmb{x} \cdot \pmb{x}}$ indicates the Euler 2-norm of a vector.  The first term on the right-hand side of \eqref{eq:geom_freq} is scalar and captures a radial dilatation, namely the rate of change of the voltage magnitude, whereas the second term is a bivector the magnitude of which embeds the information on the rotation of the voltage vector field and is proportional to the curvature of the trajectory defined by $\pmb{v}_{abc}(t)$ \cite{Milano2022_Frenet_Frame}.   Note that the magnitude of each term on the right-hand side of \eqref{eq:geom_freq} are invariants, namely, their magnitude is independent from the coordinates utilized to represent the voltage vector.

In \cite{Milano2022_Frenet_Frame}, it is shown that if the torsion of the trajectory defined by $\pmb{v}_{abc}(t)$ is zero for all $t$, then there exists a set of coordinates for which the velocity (voltage) vector can be described by only two elements, and hence the trajectory is a plane curve. In summary, the following identities hold for the quantities in \eqref{eq:fed_comp_freq}:
\begin{subequations}\label{eq:geom_equal_ICF}
\begin{equation}
    \varrho_{\mathfrak{m}}(t) = \frac{\pmb{v}_{abc}(t) \cdot \dot{\pmb{v}}_{abc}(t)}{|\pmb{v}_{abc}(t)|^2} \, ,
\end{equation}
\begin{equation}
    \omega_{\mathfrak{m}}(t) = \frac{|\pmb{v}_{abc}(t) \wedge \dot{\pmb{v}}_{abc}(t)|}{|\pmb{v}_{abc}(t)|^2} - \omega_{dq}(t) \, ,
\end{equation}
\end{subequations}
if and only if the torsion is zero, namely:
\begin{equation}
    \label{eq:torsion}
    \hat{\tau}(t) = \frac{\pmb{v}_{abc}(t) \wedge \dot{\pmb{v}}_{abc}(t) \wedge \ddot{\pmb{v}}_{abc}(t)}{|\pmb{v}_{abc}(t) \wedge \dot{\pmb{v}}_{abc}(t)|^2} = 0 \, , \quad \forall t \, .
\end{equation}
In light of the equivalence given in \eqref{eq:geom_equal_ICF}, and the interpretation of the ICF as a two-dimensional case of the geometric frequency, \textit{instantaneous planar phase} (IPP) and \textit{instantaneous planar frequency} (IPF) appear an adequate notation for the ICP and ICF, respectively, defined in \cite{Milano2022_Complex_Frequency}.  

It is possible to give a geometric interpretation also to analytic signals.  If one assumes that the original signal $v(t)$ is a velocity along a given coordinate axis, $\mathcal{H}\{v(t)\}$ can be viewed as a new coordinate along an axis orthogonal that of $v(t)$.  Then the components of the ICF are the radial and angular frequencies in the two-dimensional space of 1-D analytic signals.  In this sense, thus, Hahn's ICF is also an invariant --- and thus unique as shown in \cite{Hahn2003_Uniqueness_A_ph_analytic} ---, as the components of the geometric frequency defined in \eqref{eq:geom_freq}.  On the other hand, Milano's IPF assumes a three-dimensional space $(a, b, c)$ defined by the physical phases of the circuit.

The geometric interpretation of the Hilbert transform leads to the formal conditions that imply equivalence between Hahn's ICF and Milano's IPF. First, we observe that if \eqref{eq:torsion} holds,  there exists a plane, say $(\mu, \xi)$, where the voltage vector $\pmb{v}_{\mu\xi}(t)$ lies.  The procedure to find this plane is illustrated, for example, in \cite{Jancewicz:1989}.  If in addition to \eqref{eq:torsion}, also the following identity holds:  
\begin{equation}
    \label{eq:equiv_coordinates}
    v_{\xi}(t) = \mathcal{H}\{v_{\mu}(t)\} \, ,  \quad \forall t \, ,   
\end{equation} 
then, one obtains:
\begin{equation}
    \begin{aligned}
    \tilde{v}_{\mu}(t)
    &= \bar{v}_{\mu\xi}(t) \, , \quad \forall t \, ,        
    \end{aligned}
\end{equation}
which implies that the ICP and ICF of $\tilde{v}_{\mu}(t)$ and the the IPP and IPF of $\bar{v}_{\mu\xi}(t)$, respectively, coincide.  Equations \eqref{eq:ICP_ICF_Hahn_Milano} represent thus a special case of the geometric conditions above and corresponds to a three-phase voltage for which $(\mu, \xi) \equiv (\alpha, \beta)$ and $a \equiv \alpha$. Lastly, the suggested unified terminology is summarized in \Cref{tab:sum_table}.

\begin{table}
    \caption{Suggested Unified Terminology}
    \label{tab:sum_table}
    \centering
    \begin{tabular}{cccc}
    \toprule
    Concept & Foundation  & Definition \\
    \midrule
    Complex Frequency  & Analytic Signals & \Cref{eq:hahn_complex_freq}\\
    Planar Frequency & Space Vectors & \Cref{eq:fed_comp_freq} \\
    Geometric Frequency & Differential Geometry & \Cref{eq:geom_freq} \\
    \bottomrule
    \end{tabular}
\end{table}
\section{Conclusions}\label{sec:conclusions}
This letter provides a clarification on the concepts of ICP and ICF and identifies two coexisting but methodologically and conceptually diverse approaches that are used in current literature.  These two approaches are based on analytic signals and space vectors, respectively.  The letter provides   definitions and formulations of each approach, discusses underlying assumptions, and indicates the strict mathematical conditions under which the two approaches lead to same results.  Finally, the letter suggests an unified notation aimed at preventing confusion between these approaches.

\bibliographystyle{ieeetr}
\bibliography{references}

\end{document}